\documentstyle[12pt]{article}

\def\bq{\begin{eqnarray}}
\def\eq{\end{eqnarray}}
\def\vp{\varphi}
\def\prt{\partial}

\begin{document}
\title{\bf  An elementary quantum mechanics calculation for the 
Casimir effect in one dimension}
\author{
Attila Farkas\\
\it Institute of Condensed Matter Research, Timi\c soara,\\
\it Str. T\^ arnava 1, RO-1900 Timi\c soara, Romania\\
\\
Nistor Nicolaevici\\
\it Technical University of Timi\c soara, Department of Physics,\\
\it P-\c ta Hora\c tiu 1, RO-1900 Timi\c soara, Romania}
\maketitle

\begin{abstract}
We obtain the Casimir effect for the massless scalar field in one 
dimension based on the analogy between the quantum field and the 
continuum limit of an infinite set of coupled harmonical oscillators. 
\end{abstract}

\section{Introduction}
A well known fact in quantum mechanics is that, even though the classical 
system admits a zero minimal energy, this does not generally hold for 
its quantum counterpart. The typical example is the $\frac{1}{2}\hbar 
\omega$ value for the non-relativistic harmonic linear 
oscillator, where $\hbar$ is the Planck constant and $\omega$ its proper 
frequency. More generally, if the system behaves as a collection of such 
oscillators, the minimal (or zero point) energy is
\bq
E_0=\frac{\hbar}{2}\sum_n \omega_n\label{1},
\eq
where the sum extends over all proper frequencies $\omega_n$. As often pointed 
out in quantum field theory textbooks$^{1,2}$, non-interacting 
quantized fields can be pictured this way, in the limit of an infinite spatial 
density of oscillators. In particular, for the scalar field the analogy with a 
set of coupled oscillators can be constructed in a precise manner$^1$, as we 
shall also sketch below. We shall use here the oscillator model to obtain 
the Casimir effect for the massless field, in the case of one 
spatial dimension. The calculation is a simple exercise in non-relativistic 
quantum mechanics. 

What is usually refered to as the Casimir effect$^3$ is the attraction force 
between two conducting parallel uncharged plates in vacuum. The phenomenon 
counts as a direct evidence for the zero point energy of the quantized 
electromagnetic field: assuming the plates are perfect conductors, 
the energy to area ratio reads$^1$ ($c$ is the speed of light and $L$ is 
the plates separation) 
\bq
\frac{E_0}{A}=-\frac{\pi^2\hbar c}{720 L^3},
\label{2}
\eq
from which the attraction force can be readily derived. Qualitatively, the 
$L$ dependence in $E_0$ is naturally understood as 
originating in that displayed by the proper frequencies of the field between 
the plates.

Actually, by summing over frequencies as in eq. (\ref{1}) one obtains a 
divergent energy. This is a common situation in quantum field 
theory, being remedied by what is called 
renormalization: one basically subtracts a divergent quantity to render the result 
finite, with the justification that only energy $differences$ are 
relevant\footnote{In the assumption of neglecting gravitational phaenomena, 
see  e.g. Ref 4.}. Unfortunately, 
computational methods used to handle infinities to enforce 
this operation\footnote{i.e. regularization methods. An example follows next 
paragraph.} present themselves, rather generally, as a piece 
of technicality with no intuitive support; for the unaccustomed reader, they 
might very well leave the impression that the result is just a mathematical 
artifact. The oscillator analogy comes to provide a context to do the 
calculations within a physically transparent picture, with no extra 
mathematical input required.

\section{Quantum field theory calculation}
We briefly review first the field theoretical approach. Consider the uncharged 
massless scalar field in one dimension $-\infty<x<\infty$,
\bq
\left(
\frac{1}{c^2}\frac{\prt^2}{\prt t^2}-
\frac{\prt^2}{\prt x^2}
\right)\vp(x,t)=0\label{3},
\eq
subjected to the conditions
\bq
\vp(0,t)=\vp(L,t)=0\label{4}
\eq
for some positive $L$. We are interested in the zero point energy as a 
function of $L$. We shall focus on the field in the ``box'' $0<x<L$. It is 
intuitively clear that the result for the exterior regions follows by making $L\rightarrow \infty$. 
Note that by eqs. (\ref{4}) the field in the box is causally 
disconnected from that in the exterior regions, paralleling thus the situation 
for the electromagnetic field in the previous chapter.

Eqs. (\ref{3}) and (\ref{4}) define the proper frequencies as
\bq
\omega_n=\frac{n\pi}{L},
\quad n=1,2,\dots\infty,
\eq
obviously making $E_0$ a divergent quantity. A convenient way$^5$ to deal 
with this is by introducing the damping factors
\bq
\omega_n \rightarrow \omega_n \exp (-\lambda \omega_n/c),
\quad \lambda>0,
\eq
and to consider $E_0=E_0(L,\lambda)$ in the limit $\lambda\rightarrow 0$. 
Performing the sum one obtains
\bq
E_0(L,\lambda)=\frac{\pi \hbar c}{8L}\left(\mbox{cth}^2\frac{\pi \lambda}{2L}-1
\right).
\eq
Using the expansion
\bq
\mbox{cth}\, z=\frac{1}{z}+\frac{z}{3}+{\cal O}(z^3),\label{exp1}
\eq
one finds
\bq
E_0(L,\lambda)=\frac{\hbar c}{2\pi\lambda^2}L
-\frac{\pi\hbar c}{24L}
+{\cal O}\left(\frac{\lambda}{L}\right).
\label{as1}
\eq
Now, it is immediate to see that the $\lambda^{-2}$ term can be assigned to an 
infinite energy density corresponding to the case $L\rightarrow \infty$. 
The simple but essential observation is that, when considering also 
the energy of the exterior regions, the divergences add to an $L$-independent 
quantity, which makes them mechanically irrelevant. Renormalization amounts to 
ignore them. Thus one can set
\bq
E_0(L)=-\frac{\pi}{24}\frac{\hbar c}{L},
\label{e0}
\eq
which stands as the analogous result of eq. (\ref{2}).

\section{Quantum mechanics calculation}
Consider the one dimensional system of an infinite number of coupled oscillators 
described by the Hamiltonian (all notations are conventional)
\bq
H=\sum_k \frac{p_k^2}{2m}+\sum_k\frac{k}{2}(x_{k+1}-x_k)^2.
\label{h}
\eq
$x_k$ measures the displacement of the $k$th oscillator from its 
equilibrium position, supposed equally spaced from the neighbored ones 
by distance $a$. Canonical commutations assure that the Heisenberg 
operators
\bq
x_k(t)=e^{\frac{i}{\hbar}Ht}x_k e^{-\frac{i}{\hbar}Ht}
\eq
obey the classical equation
\bq
m\frac{d^2 x_k(t)}{dt^2}-k(x_{k+1}(t)+x_{k-1}(t)-2x_k(t))=0.
\label{classic}
\eq
Let us consider the parameters $m$ and $k$ scaled such that 
\bq
a^2\frac{m}{k}=\frac{1}{c^2}.
\eq
As familiar from wave propagation theory in elastic media, eq. (\ref{classic}) 
becomes the d'Alembert equation (\ref{3}) with the correspondence 
\bq
x_k (t)\rightarrow\vp(k a,t),
\label{ana}
\eq 
and letting $a\rightarrow 0$. $x_k$, $p_m$ commutations can be also shown 
to translate into the equal-time field variables commutations required by 
canonical quantization$^1$. One can thus identify 
the quantum field with the continuum limit of the quantum mechanical 
system.

Our interest lies in the oscillator analogy when taking into account 
conditions (\ref{4}). It is transparent from eq. (\ref{ana}) that they 
formally amount to set in $H$
\bq
x_0=x_N=0,\quad p_0=p_N=0,
\eq
with $N$ some natural number. In other words, the 0th and the $N$th 
oscillator are supposed fixed. As in the precedent paragraph, we shall 
calculate the zero point energy of the oscillators in the ``box'' 
$1\leq k \leq N-1.$

The first step is to decouple the oscillators by diagonalizing the quadratical 
form in coordinates in eq. (\ref{h}). Equivalently, one needs 
the eigenvalues $\lambda_n$ of the $N-1$ dimensional square matrix $V_{km}$ 
with elements
\bq
V_{k,k}=2,\quad V_{k,k+1}=V_{k,k-1}=-1,
\eq
and zero in rest. One easily checks they are
\bq
\lambda_n=4\sin^2\frac{n\pi}{N},\quad n=1,2,\dots N-1,
\eq
with $\lambda_n$ corresponding to the (unnormalized) eigenvectors 
$x_{n,k}=\sin\frac{nk}{N}$. It follows
\bq
E_0(N,a)=\frac{\hbar c}{a}\sum_{n=1}^{N-1}\sin\frac{n\pi}{2N}.
\label{en}
\eq
To make connection with the continuous picture, we assign to the system the 
length 
\bq
L=aN
\label{N}
\eq
measuring the distance between the fixed oscillators, and eliminate $N$ in 
favour of $a$ and $L$ in eq. (\ref{en}). After summing the series one obtains
\bq
E_0(L,a)=\frac{\hbar c}{2a}\left(
\mbox{ctg} \frac{\pi a}{4L}-1\right).
\eq
With an expansion similar to eq. (\ref{exp1})
\bq
\mbox{ctg}\, z=\frac{1}{z}-\frac{z}{3}+{\cal O}(z^3),
\label{exp}
\eq
it follows for $a\ll L$
\bq
E_0(L,a)=\left(\frac{2\hbar c L}{\pi a^2}-\frac{\hbar c}{2a}\right)-
\frac{\pi}{24}\frac{\hbar c}{L}+{\cal O}\left(\frac{a}{L}\right).
\label{a2}
\eq

The result is essentially the same with that in eq. (\ref{as1}). 
The $a$ independent term reproduces the renormalized value (\ref{e0}). 
An identical comment applies to the $a\rightarrow 0$ diverging terms. Note 
that the $L\rightarrow \infty$ energy density can be equally obtained 
by making $N\rightarrow \infty$ in eq. (\ref{en}) and evaluating 
the sum as an integral. Physically put, this corresponds to an infinite crystal with 
vibration modes characterized by a continuous quasimomentum in the Brillouin 
zone
\bq
0\leq k<\frac{\pi}{a},
\eq
and dispersion relation
\bq
\omega(k)=\frac{2 c}{a}\sin \frac{ka}{2}.
\eq
Note also that the second term, with no correspondent 
in eq. (\ref{as1}), can be absorbed into the first one with an irrelevant 
readjustment of the box length $L\rightarrow L-\frac{\pi a}{4}$.

\section{Quantum field vs oscillator model: quantitative comparison 
and a speculation}

Let us define for $a>0$ the subtracted energy $E_0^S(L,a)$ as the difference 
between $E_0(L,a)$ and the paranthesis in eq. (\ref{a2}), so that
\bq 
\lim_{a\rightarrow 0} E^S_0(L,a)=E_0(L).
\eq 
One may ask when the oscillator 
model provides a good approximation for the quantum field, in the sense that
\bq
\frac{E^S_0(L,a)}{E_0(L)}=-3\left \lbrace\left(\frac{4L}{\pi a}\right) \mbox{ctg} 
\left (\frac{\pi a}{4L}\right) -\left(\frac{4L}{\pi a}\right)^2 \right\rbrace
\label{as}
\eq
is close to unity. Note that by eq. (\ref{N}) expression above is a function of $N$ 
only. The corresponding dependence is plotted in Fig.1. One sees, quite 
surprisingly, that already a number of around twenty oscillators suffices to 
assure a relative difference smaller than $10^{-4}$. More precisely, one has 
that the curve assymptotically approaches zero as 
\bq
\frac{\pi^2}{240} \frac{1}{N^2}.
\label{asym}
\eq

We end with a bit of speculation. Suppose there exists some privileged 
scale $l$ (say, the Plank scale) which imposes a universal bound for 
lengths measurements, and consider the oscillator system with the spacing given 
by $l$. The indeterminacy in $L$ will cause an indeterminacy in energy 
(we assume $L\gg l$)
\bq
\frac{\Delta E_0^S}{E^S_0}\sim
\frac{\Delta E_0}{E_0}\sim\frac{l}{L}.
\eq
On the other hand, the assymptotic expression (\ref{asym}) implies 
\bq
\frac{E_0^S-E_0}{E_0}\sim\left(\frac{l}{L}\right)^2.
\eq
We are led thus to the conclusion that, as far as Casimir effect measurements 
are considered, one could not distinguish between the ``real'' quantum 
field and its oscillator model.

\end{document}